\documentclass{PoS}
\usepackage{bm}

\title{Opportunities for spin physics at EIC}

\ShortTitle{Spin Physics Opportunities at EIC}

\author{\speaker{Dani\"el Boer}\thanks{I thank Ernst Sichtermann for very useful feedback.}\\
        Van Swinderen Institute for Particle Physics and Gravity, University of Groningen\\ 
Nijenborgh 4, NL-9747 AG Groningen, The Netherlands\\
        E-mail: \email{d.boer@rug.nl}}

\abstract{This is a brief overview of the spin physics opportunities at a high energy, high luminosity, polarized Electron-Ion Collider (EIC).
It covers measurements of electroweak polarized structure functions, quark and gluon PDFs, TMDs, GPDs and GTMDs. Exploiting the many possible final states allows to probe various spin effects. Open and bound heavy quark production can be used to probe gluon TMDs, but also color-octet NRQCD long distance matrix elements. Spin-dependent fragmentation functions can be used too, but are also interesting in themselves. Especially for studies of the small-$x$ and the high-$Q^2$ spin structure the EIC will be essential. 
}

\FullConference{XXVII International Workshop on Deep-Inelastic Scattering and Related Subjects - DIS2019\\
		8-12 April, 2019\\
		Torino, Italy}

\begin{document}

\section{Introduction}
The EIC is a high-energy, high-luminosity electron-ion collider with the capability to accelerate polarized hadrons. It offers many opportunities to study spin effects. Some have already been investigated experimentally but can be extended at the EIC, others have never been performed before, even if some of the suggestions have been made already decades ago. Many of the proposed and promising spin studies have been outlined in the EIC white paper \cite{Accardi:2012qut} (see also \cite{Boer:2011fh}), but with the growing prospect of the EIC's realization, additional suggestions have been made since. In this brief overview these will be described, starting with the one-dimensional spin structure of protons, neutrons, and deuterons, followed by a discussion of the three-dimensional spin structure in terms of transverse momentum dependent parton distributions (TMDs), which includes the Sivers effect and a gluon polarization effect in unpolarized collisions. Subsequently, the transverse spatial distributions as given by the Generalized Parton Distributions (GPDs) and Generalized TMDs (GTMDs) are discussed, followed by a final section on fragmentation functions as analyzers of spin structure.

\section{One-dimensional spin structure}
The one-dimensional spin structure of hadrons is expressed in terms of structure functions and collinear parton distribution functions (PDFs).
In earlier polarized Deep Inelastic Scattering (DIS) experiments the polarized structure functions $g_1$ and $g_2$ have been extracted. In the pQCD improved parton model $g_1$ provides information about the contributions of the quark, antiquarks and gluons to the proton spin through the helicity PDFs $\Delta q(x,Q^2)$ and $\Delta g(x,Q^2)$. The EIC will extend the range in $x$ allowing for a reduction in the uncertainty on the integrated quantities entering the spin sum rule, in particular on $\Delta g(Q^2) = \int_0^1 dx \Delta g(x,Q^2)$. The structure function $g_2$ is of interest because of the Burkhardt-Cottingham sum rule, $\int_0^1 dx\ g_2(x,Q^2) = 0$, and because of its twist-3 part which describes the deviation from the Wandzura-Wilczek approximation. The second Mellin moment of this twist-3 part, $d_2 = 3 \int_0^1 dx \left. x^2 \, g_2(x,Q^2)\right|_{\rm twist-3}$, turns out to be small \cite{Armstrong:2018xgk}, but is of interest because it can be compared to its lattice evaluation. Furthermore, mapping out the $Q^2$-dependence of $d_2$ is interesting in order to test the expectations about the evolution of twist-3 functions, see e.g.\ \cite{Braun:2011aw}.

At high $Q^2$ (and correspondingly high $x$ values) one can become sensitive to three additional structure functions, which require weak interactions. They are generally called $g_3$, $g_4$, and $g_5$, but with sometimes differing definitions, cf.\ \cite{Blumlein:1996vs}. Charged current DIS allows to extract information about the charm helicity distribution $\Delta c$ from the comparison of $g_1^{W^-}$ and $g_5^{W^-}$, cf.\ e.g.\ \cite{Zhao:2016rfu}. 

At the EIC there are possibilities to access the spin structure of the neutron, which is interesting in comparison to polarized protons, for instance through the Bjorken sum rule. Polarized deuterons would allow to probe polarized neutrons through spectator tagging in the process $ed \to e'pX$ \cite{Cosyn:2016oyw}. Another option is to use helium-3 polarization which to a large extent arises from the neutron. 

Polarized deuterons would furthermore allow to probe four additional structure functions: $b_1$-$b_4$ \cite{Hoodbhoy:1988am}. Thus far only $b_1$ has been extracted \cite{Airapetian:2005cb}. It requires longitudinal tensor polarization of the deuteron, but no lepton polarization. The measured $b_1$ becomes nonzero for $x$ values below 0.1 and increases in magnitude towards smaller $x$. It would be interesting to measure $b_1$ more precisely and over a wider $x$ range at the EIC, because it gives information about the partonic contributions to the strong force beyond the confinement range. After all, the deuteron is a very loosely bound state of a proton and a neutron and $b_1$ is absent for protons and neutrons separately. 

A new polarized deuteron measurement that might be possible at the EIC is that of the transverse tensor polarization distribution $h_{1TT}$ (in the notation of \cite{Bacchetta:2000jk}). It does not arise in the quark parton model. It is only nonzero for gluons and arises at leading twist \cite{Jaffe:1989xy,Artru:1989zv}. 
    
\section{Three-dimensional momentum space spin structure}
TMDs provide information about the three-dimensional momentum space structure in $x$ and transverse momentum $\bm{k}_T$. Quark TMDs can be probed in for instance semi-inclusive DIS (SIDIS), where a final state hadron $h$ is observed ($ep \to e'\, h \, X$). An important objective of EIC is to measure the Sivers effect that correlates the transverse momentum with the transverse spin of the proton. The Sivers effect in SIDIS has been clearly observed by the HERMES and COMPASS experiments \cite{Airapetian:2009ae,Alekseev:2010rw}. Due to the higher energy, at the EIC the quark Sivers effect can be measured in SIDIS with an observed final state jet: $e \, {p}^\uparrow \to e' \, {\rm jet} \, X$. In the approximation where the jet has an infinitely narrow transverse momentum distribution, the single spin asymmetry is given by:
\begin{equation}
A_{UT}^{\sin(\phi_{\small {\rm jet}}^e-\phi_{S}^e)} \propto\;|\bm S_{T}^{}|
\; \frac{Q_T}{M} {f_{1T}^{\perp q}(x,Q_T^2)}, \qquad
Q_T^2 = |\bm P^{{\rm jet}}_{\perp}|^2, 
\end{equation}
where the transverse momentum of the quark Sivers function $f_{1T}^{\perp q}$ can be accessed directly (in reality with some smearing). Besides this advantage, at the EIC it can be measured in the same kinematic range as in Drell-Yan (DY) experiments, allowing for a test of the overall sign change relation, $f_{1T}^{\perp q [{\rm SIDIS}]}(x,k_T^2) = - f_{1T}^{\perp q [{\rm DY}]}(x,k_T^2)$ \cite{Collins:2002kn}, without the additional theoretical uncertainty from TMD evolution.
Measurements of the Sivers effect in DY are part of ongoing programs at CERN, Fermilab, BNL and future programs, e.g.\ at NICA in Dubna. The first data on DY from COMPASS \cite{Adamczyk:2015gyk} and on $W$-boson production from STAR \cite{Aghasyan:2017jop} are compatible with the sign-change. 

The gluon Sivers effect can among others be probed in open heavy quark pair production ($e\, p^\uparrow \to e' \, Q\, \overline{Q}\, X$), such as $D$-meson pair production ($e\, p^\uparrow \to e' \, D\, \overline{D}\, X$), where the transverse momentum of the pair is observed.
It enters in the following sign change relation \cite{Boer:2016fqd}:
\begin{equation}
f_{1T}^{\perp\, g \, [e\, p^\uparrow \to e' \, Q\, \overline{Q}\, X]}(x,p_T^2) = - f_{1T}^{\perp\, g \, [p^\uparrow\,  p\to \gamma \, \gamma \, X]} (x,p_T^2).
\end{equation}
Measuring di-photon production at RHIC will be very challenging, as will open heavy quark pair production at the EIC. The Sivers asymmetry in the latter is given (at leading order) by \cite{Boer:2016fqd}
\begin{equation}
A_{UT}^{\sin(\phi_S-\phi_{\scriptscriptstyle T})} =  \frac{\vert \bm q_{\scriptscriptstyle T}\vert}{M_p}\, \frac{f_{1T}^{\perp\,g}(x,\bm q_{\scriptscriptstyle T}^2)}{f_1^g(x,\bm q_{\scriptscriptstyle T}^2)}.
\end{equation}
This asymmetry is maximally 1, but in reality it is expected to be smaller, of course. For a Sivers function that is 10\% of what is maximally allowed by the positivity constraint, the asymmetry cannot be discerned within the statistical uncertainty at the EIC, assuming a realistic $L_{\rm int} = 10\ {\rm fb}^{-1}$ \cite{Zheng:2018awe}. The asymmetry in dijet production offers a better chance, but the additional contribution from the quarks makes this a less clean observable \cite{Boer:2016fqd}.

Another opportunity at the EIC is to measure the distribution $h_1^{\perp\; g}$ of linearly polarized gluons in unpolarized hadrons through the process of open heavy quark pair production in both $ep$ and $eA$ collisions. This involves for instance \cite{Boer:2010zf} the measurement of a $\cos 2(\phi_{\scriptscriptstyle T} - \phi_\perp)$ distribution, where $\phi_{\scriptscriptstyle T}$ and $\phi_\perp$ are the angles of the sum and the difference of the transverse momenta of the two heavy quarks, respectively. The bound on this angular asymmetry can reach 15\% for both charm and bottom quarks \cite{Pisano:2013cya}. Unlike the Sivers function, the function $h_1^{\perp\; g}$ may actually be maximal or close to it, especially for small $x$ and larger tranverse momenta \cite{Metz:2011wb}. It is expected to keep up with the growth of the unpolarized gluon distribution as $x \to 0$. In the small-$x$ McLerran-Venugopalan model asymmetries on the 10\% level can be attained especially for larger $Q^2$ values \cite{Boer:2016fqd}. Similarly results are found for dijet production \cite{Pisano:2013cya,Dumitru:2015gaa,Boer:2016fqd}. The angular asymmetry is expected to have opposite signs for $L$ and $T$ photon polarizations, which will be interesting to test at the EIC \cite{Dumitru:2018kuw}.
  
Polarized TMDs can also be accessed in quarkonium production, i.e.\ $e \, p^\uparrow \to e' \, {[Q\overline{Q}]} \, X$ where $[Q\overline{Q}]$ denotes a bound quarkonium state, like a $J/\psi$ or $\Upsilon$ \cite{Godbole:2012bx,Godbole:2013bca,Godbole:2014tha,Mukherjee:2016qxa,Rajesh:2018qks}. In leading order (LO) the quarkonium is a color octet (CO), which in a combined TMD and NRQCD approach means that the spin asymmetries depend on long distance matrix elements (LDMEs) that are quite uncertain. Ratios of asymmetries allow to cancel those out at LO \cite{Bacchetta:2018ivt}, offering ways to access the various gluon TMDs. Conversely, by comparing to the process of open heavy quark pair production $e \, p \to e' \, Q \, \overline{Q} \, X$ one can construct ratios in which the TMDs cancel out at LO, offering a new way to extract the CO NRQCD LDMEs. For a discussion of the robustness of this result and of how to use the polarization of the quarkonia to validate the extractions, see \cite{Bacchetta:2018ivt}.
  
\section{GPDs and GTMDs}

Information about the spatial distribution of quark and gluons inside protons or nuclei can be obtained at the EIC as well. 
Quark GPDs for longitudinally polarized protons will be extracted in order to shed light on the quark orbital angular momentum component in the proton spin sum rule. Helicity flip or transversity GPDs as well Sivers-like distortions in coordinate space, i.e.\ $b_T \times S_T$ correlations related to the derivative of the GPD $E$, can be studied using transverse spin asymmetries. 
Generalized TMDs (GTMDs) can be viewed as transverse momentum dependent GPDs or alternatively as off-forward TMDs: $G(x,\bm{k}_T,\bm{\Delta}_T)$. They can also be viewed as Fourier transforms of Wigner distributions $W(x,\bm{k}_T,\bm{b}_T)$ \cite{Ji:2003ak,Belitsky:2003nz,Meissner:2009ww}. The so-called elliptic gluon GTMD ($\propto \cos 2 \phi_{\bm{k} \bm{\Delta}}$) \cite{Hatta:2016dxp,Zhou:2016rnt,Boer:2018vdi} is of particular interest at small $x$. The description of DVCS at small $x$ requires inclusion of its Fourier transform, the elliptic Wigner function, that contributes to the transversity gluon GPD $E_T$ \cite{Hatta:2017cte}. The first suggestion to measure gluon GTMDs through hard diffractive dijet production \cite{Altinoluk:2015dpi,Hatta:2016dxp} can, although probably very challenging, be studied at the EIC.   

\section{Fragmentation functions as analyzers of spin distributions} 

Fragmentation functions (FFs), both collinear and transverse momentum dependent ones, can be used to do spin physics at the EIC using the lighter quarks in the final state ($u,d,s$). The  Collins effect FF TMD and the collinear chiral-odd two-hadron FF allow to probe quark transversity distribution $h_1^q$ \cite{Collins:1993kq,Collins:1994ax,Jaffe:1997hf,Bianconi:1999cd,Radici:2001na,Radici:2018iag}. Two-hadron FFs also allow to probe the quark helicity distributions $g_1^q=\Delta q$, but that requires inclusion of transverse momentum dependence. This involves the `handedness' FF $G_1^\perp$ \cite{Bianconi:1999cd}, which is around 10\% of the size of the unpolarized FF $D_1$ according to the model of \cite{Matevosyan:2018oui}. See \cite{Matevosyan:2017liq,Matevosyan:2018icf} for details on its possible experimental extraction from $e^+ e^-$ data.  
  
Using polarized $\Lambda$'s $g_1^q$ can be probed through the polarization transfer $D_{LL}$ \cite{Belostotski:2011zza,Alekseev:2009ab,Adam:2018kzl}, which moreover is interesting for studying the spin sum rule for $\Lambda$'s. Similarly $h_1^q$ can be probed through $D_{NN}$, although it appears to be small \cite{Negrini:2009oia,Adam:2018wce}. 

Polarized $\Lambda$'s are also produced in unpolarized collisions, as is well established in $pA$ collisions, but not yet in $eA$ collisions. The only SIDIS data in the current fragmentation region are from NOMAD \cite{Astier:2000ax} and ZEUS \cite{Chekanov:2006wz} and both are compatible with zero within large errors. Measurements of $\Lambda$ polarization in SIDIS could clarify the underlying mechanism which in the TMD formalism is described in terms of $D_{1T}^\perp$, the so-called polarizing fragmentation function \cite{Mulders:1995dh,Anselmino:2001js}. The latter can be extracted from $e^+ e^-$ experiments \cite{Boer:1997mf} and recent data by the BELLE Collaboration \cite{Abdesselam:2016nym,Guan:2018ckx} clearly show it to be nonzero, providing additional motivation to measure polarized $\Lambda$'s at the EIC.

\end{document}